# Using Context Dependent Semantic Similarity to Browse Information Resources:
# an Application for the Industrial Design

Riccardo Albertoni, Monica De Martino

CNR-IMATI,
Via De Marini, 6 – Torre di Francia - 16149 Genova, Italy
{albertoni, demartino}@ge.imati.cnr.it

**Abstract.** This paper deals with the semantic interpretation of information resources (e.g., images, videos, 3D models). We present a case study of an approach based on semantic and context dependent similarity applied to the industrial design. Different application contexts are considered and modelled to browse a repository of 3D digital objects according to different perspectives. The paper briefly summarises the basic concepts behind the semantic similarity approach and illustrates its application and results.

## 1 Introduction

An increasing number of activities in several disciplinary and industrial fields such as scientific research, industrial design, collaborative design and environmental management, rely on the production and employment of information resources representing objects, information and knowledge. The vast availability of digitalized information resources (documents, images, videos, 3D model) highlights the need for appropriate methods to effectively share and employ all of them. In particular, tools to search for information resources produced by third parties are required to successfully achieve our daily work activities. The search for information resources is a painstaking activity. To decide which are the most suitable resources for specific and severely constricted tasks requires a careful comparison of the available information resources. A "Google-like" list of ranked items and some web pages excerpts are probably enough for a typical internet seeker, but they are not sufficient to decide which resources best fit a given task as for example in the case of a designer.
Headway in this direction is made adopting metadata and ontology. Metadata are a description of the most relevant features characterizing the information resources. Ontology represents an explicitly shared conceptualisation of a specific domain knowledge that a community has agreed upon: it consists of different parts, including a set of concepts and their mutual relations and instances related to the domain knowledge to be represented. Knowledge conceptualization using ontologies provides means to organize the *metadata* (called ontology-driven metadata) of complex information resources. According to Sheth et al. [1] ontology-driven metadata provide syntactic and semantic information about information resources. *Syntactic metadata* describe non-contextual information about the content (e.g. language, a bit rate, format). This offers no insight into the meaning of a document. In contrast, *semantic metadata* provide the content and contextual information of a specific domain, such as which entities take part in the production and usage of the information resource. The metadata of the resources are encoded as instances in the ontology. However, a plenty of features have to be considered to fully describe the information resources in sophisticated fields as those mentioned. This brings to a complexity of metadata and to a growing need for tools facing with this complexity and their analysis.

The paper deals with the problem of metadata complexity proposing the use of semantic similarity for the browsing of resources by their ontology-driven metadata. Semantic similarity plays an important role in information systems as it supports the identification of objects that are conceptually close but not identical. We have recognised the importance of solving this problem from our direct research experience in the AIM@SHAPE [2] initiative. It is a network of excellence, where ontologies have been adopted to organize the metadata of complex information resources: different ontologies have been defined to describe 3D/2D models (e.g. models of mechanical objects, digital terrains or artefacts from cultural heritage), tools for processing the models and implicit knowledge about the functionality embedded in a 3D shape [3], [4], [5], [6]. From our experience, the ontology driven metadata definition turns out to be outrageously expansive in terms of man-month efforts needed, especially whenever the domain that has to be formalized is complex and compound. The "standard ontology technology" provides reasoning facilities that are very useful in supporting querying activity as well as in checking ontology consistency, but the current technology lacks an effective tool for comparing the resources (instances). Although, the efforts necessary to formalize the ontology, domain experts are often quite willing to provide the domain knowledge required to characterize their resources. However, they are disappointed when their efforts do not result in any measure of similarity among the resources. Aware of this shortcoming, we have addressed our research efforts on how to better employ the information encoded in the ontology and to provide tools that exploit as much as possible the result of the aforementioned efforts [7], [8]. In [8], [9] we have illustrated an innovative approach to assess a context dependent semantic similarity among ontology instances. We refer to the assessment of the similarity as "semantic" because our approach exploits the "semantics" of information resources in term of their resources origin, important features and potential usages formalised in the ontology driven metadata.

In this paper we aim to illustrate the usefulness of our semantic similarity [8] [9] to browse a repository of industrial design models. We suppose that a designer has to browse a large repository of 3D objects and we assume that he/she has to search for similar objects representing "kitchen accessories" in particular "containers". For this scope, we have defined a simple ontology, which formalises the main object properties and characteristics. Metadata describing some "containers" provided by Alessi collection are considered as instances of the ontology. We have applied our asymmetric and context dependent similarity approach considering two different application contexts. From the analysis of the results, we illustrate the contributions of the application of our approach in an industrial design activity. In particular, the

"context dependence" aids to get a designer-oriented browsing: the designer can formulate, learn and modify the criteria of similarity induced by the application context. The asymmetry is helpful to compare the objects' complexity providing hints for the re-use of existing objects. Moreover, the case study shows examples where the semantic-based approach is able to recognize similarities that can not be identified by geometric/structural based approaches.

The paper is organized as follows: in the first section we provide a short description of our approach for the semantic similarity assessment, in the next sections we describe its application for the industrial design. First we illustrate the ontology model utilized for 3D digital objects and two scenarios of reference with the related application contexts are delineated. The assessment of the semantic similarity and the analysis of the results follow. Conclusion and future extension of the work end the paper.

## 2 Semantic similarity approach

In this paragraph, we briefly provide the description and the main concepts of the approach to measure semantic similarity applied to browse 3D objects by their ontology-driven metadata.

So far, the most of research activity pertaining to similarity and ontologies has been carried out within the field of ontology alignment [10][11] or to assess the similarity among concepts [12][13]. Unfortunately, all these methods result inappropriate for the similarity among instances. On the one hand, the similarities for the ontology alignment strongly focus on the comparison of the structural parts of distinct ontologies, therefore their application to assess the similarity among instances might result misleading. On the other hand, the concepts' similarities mainly deal with lexicographic ontologies ignoring the comparison of the instances values. Apart from them, few methods to assess similarities among instances have been proposed [14][15]. Unfortunately, these methods rarely take into account the different hints hidden in the ontology and they do not consider that the ontology entities differently concur in the similarity assessment according to the application context. Some studies combine context and similarity [16][18][17]: they formalise the context taking into account the features that are relevant but ignoring the need of operations in order to properly compare them. Moreover, they do not directly address the similarity among instances.

In [8] [9] we have defined the approach, which is used in this paper, aiming at providing a new sensitive measurement of semantic similarity among the instances of the ontology exploiting as much as possible the information encoded in the ontology. Moreover, it is sensible to specific contexts inasmuch as different contexts induce different criteria of similarity.

The similarity is asymmetric in order to point out the concept of "containment" defined by: "Given two resources x, y (represented as instances in the ontology) and their sets of characteristics (coded as instance attributes and relation values), x is contained in y if the set of characteristics of x is contained[1] in the set of characteristics of y". Thus given two instances x, y, their similarity is sim(x,y)=1 if and only if the set of characteristics of x is contained in the set of characteristics of y. On the contrary, unless y is contained in x, the similarity between y and x is sim(y,x)<1. The similarity value between x and y tends to decrease as long as the level of containment of their sets of properties decreases. Of course, the containment has to consider also the inheritance between the classes: if x belongs to a sub-class of the class of y, the asymmetric evaluation is performed relying on the idea that humans perceive similarity between a sub-concept and its super-concept as greater than the similarity between the super-concept and the sub-concept [18].

The similarity assessment is structured in terms of *data, ontology* and *context* layers plus the *domain knowledge* layer, which spans all the others according to the framework proposed as adopted by Ehrig et.al. [19]. The *data layer* measures the similarity of entities by considering the data values of simple or complex data types such as integers and strings. The *ontology layer* considers the similarities induced by the ontology entities and the way they are related to each other. The *context layer* assesses the similarity according to how the entities of the ontology are used in some external contexts. In particular, we have extended this view of framework defining a *context layer*, which includes an accurate formalization of the criteria of similarity in order to tailor the similarity evaluation with respect to a context and in the definition of an *ontology layer* explicitly parameterized according to these criteria.

The context layer has a crucial role in the similarity assessment as we have demonstrated in [8] and [9] the criteria of similarity are context-dependent: the context affects the choice of entities (classes, attributes and relations) that are considered in the semantic similarity assessment as well as the different operations, which can be used to compare them. Concerning the operations our approach assumes but it is not limited to three operations (named *Count, Inter* and *Simil*) which correspond to compare the set of attributes or relations respectively according to their cardinality, their intersection and their similarity.

In this paper we adopt the formalization of application context illustrated in [8][9] to express the similarity criteria. The formalization relies on the concept of a "sequence of elements belonging to a set X", which formalizes generic sequences of elements, and a "path of recursion of length i" to track the recursion during the similarity assessment. It represents the recursion in term of the sequence of relations used to navigate the ontology. The *application context* is a function defined inductively according to the length of the path of recursion. It yields the set of attributes and relations as well as the operations to be used computing the semantic similarity.

**Definition 1: Application context AC** *Given the set P of paths of recursion, $L = \{Count, Inter, Simil\}$ the set of operations adopted, A the set of attribute in the ontology, R the set of relation, an application context is defined by a partial function AC having the signature $AC: P \rightarrow (2^{A \times L}) \times (2^{R \times L})$, yielding the attributes and relations as well as the operations to perform their comparison.*

---
[1] The containment is not meant as proper containment. In other words each set A is considered as an A subset.

A more formal definition and a complete discussion concerning the rationale behind the application context design has been provided in [8].

The formalization of the application context is employed to parameterize the computation of the similarity in the *ontology layer*, forcing it to adhere to the criteria induced by the context.

The ontology layer provides the measurement of semantic similarity. It is represented by an amalgamation function, which aggregates different similarity measurements considering hints lying at different levels such as the structural comparison between two instances in terms of the classes that the instances belong to, and the instances comparison in term of their attributes and relations. In particular, it is characterised by two similarity functions named *external similarity* and *extensional similarity*.

The *external similarity* performs a structural comparison between two instances $i_1, i_2$ in terms of the classes $c_1$, $c_2$ the instances belong to. It consists of two similarity evaluations:

- Class Matching, which is based on the distance between the classes $c_1$, $c_2$ and their depth respect to the class hierarchy induced by IS-A relationship.
- Slot Matching, which is based on the number of attributes and relations shared by the classes $c_1$, $c_2$ with respect to the overall number of their attributes and relations. Then two classes having a plenty of attributes/relations, some of whose are in common, are less similar than two classes having less attributes but the same number of common attributes/relations.

The *extensional similarity* performs the instances comparison in term of their attributes and relations. Its evaluation is parametric with respect to the assessment criteria induced by the application context explicitly formalized at the context layer. Through this formalization, it is possible to tailor the similarity to specific application need.

The similarity algorithm is developed in JAVA relying on the API provided by OWL-Protégé.

## 3 Case study for industrial design

This paragraph presents an exemplificative case study to show the worth of our context-dependent semantic similarity in order to support a user oriented browsing of multimedia contents. We refer to a case study for the industrial design: we suppose that a designer has to browse a large repository of objects in order to avoid the re-design of similar objects or for the re-using of similar design model. Each object in the repository is supposed to be described by its own ontology-driven metadata and the designer has to analyse the metadata to browse the repository.

Below, in a first section we describe a simple ontology schema to describe industrial design models, in the second section we identify two scenarios which require different criteria of similarity to compare the models and we formalize the application contexts to parameterize the similarity algorithm. In the third section, we apply the algorithm to a set of objects and we provide an analysis of the obtained results.

### 3.1 Ontology schema for kitchen containers

For the case study we have defined the domain that the ontology has to formalise: in our case, it is for the search of kitchen accessories in particular for containers.

We have defined a very simple ontology schema to describe container models. It is informally depicted in Figure 1. Notice that a full characterization of the metadata pertaining to industrial design models is out the purpose of this paper. Here the ontology is intended as a simple example to show the potentialities of our similarity approach.

The ontology provides some information about a product model such as its subdivision into components (e.g., handle, body, filter), its shape representation (e.g., mesh, graph), the tasks an object is designed to accomplish with (e.g., to cook, to boil), the functionalities provided by its components (e.g, to pour, to contain, to sieve). The ontology is specified in a subset of Ontology Web Language[2].

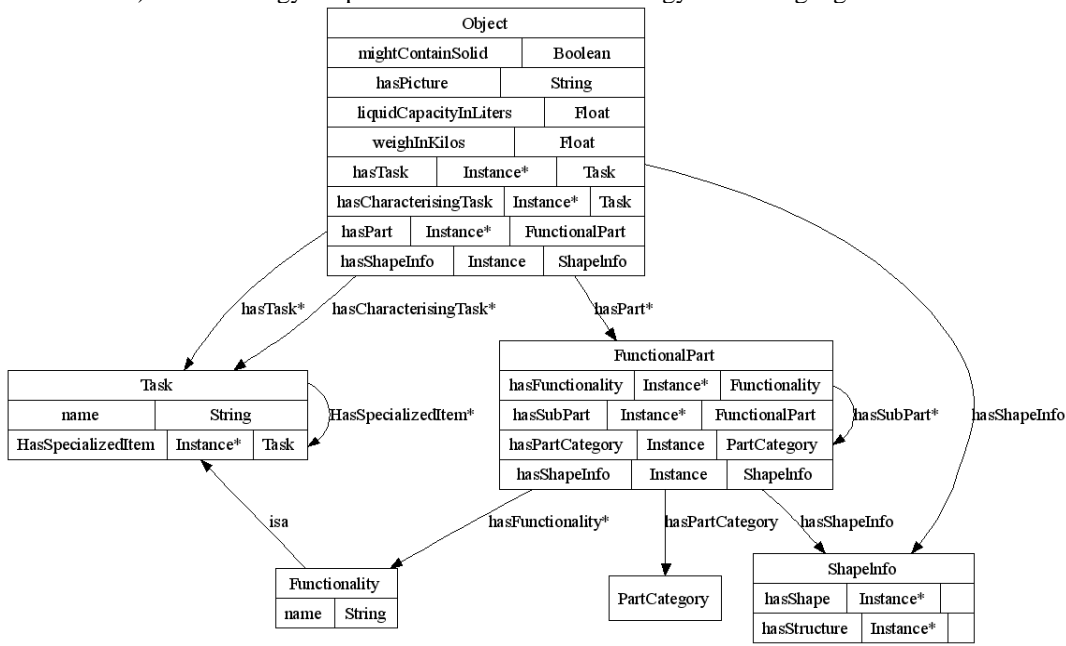

**Figure 1: Ontology schema to model metadata of objects in industrial design repository.**

The main classes in the ontology are:
- *Object*: each instance of this class represents a "container" available in the repository. The following information are included as metadata: the weight of the real object (attribute *weightInKilos*), its primary usage (relation *hasCharacterizingTask*), its decomposition in functional subparts (relation *hasPart*). Moreover, it includes also a reference to its 3D shape information (relation *hasShapeInfo*) and a URL reference to its picture (attribute *hasPicture*).

---
[2] http://www.w3.org/2004/OWL/

- *Task:* it represents the possible usages of an object (e.g. toBoil, toHeat, toDrink). Its dependences to subtasks and functionalities can be tracked by the relation *hasSpecializedItem*.
- *FunctionalPart:* it provides information about the functional parts, which compose the objects (e.g., Cover, Handle, LiquidProofConcavity, Spout, SupportingBase). Each instance of *FunctionalPart* is described by: the functionalities of the part (relation *hasFunctionality*), a reference to its subparts (relation *hasSubPart*), the category it belongs to (relation *hasPartCategory*) and a reference to the information pertaining to its 3D Shape (relation *hasShapeInfo*).
- *Functionality:* it represents special kinds of tasks, which are provided by the functional parts. On the contrary to the information provided by the class *Task*, it represents basic functionalities (e.g. ToCover, toPour, ToStabilize) usually provided by object parts (e.g. Covers, Lips, Supporting Bases). They are identified by the attribute *name* and can be structured in sub-functionalities by the relation *hasSpecializedItem*.
- *ShapeInfo:* it contains information pertaining to digital shapes representing an object or its parts. The relation *hasShape* refers to the 3D geometric model. The relation *hasStructure* refers to information about the structure (e.g. Reeb's Graph, Skeleton, Medial Axis, Segmented Parts [19]).

In this case study, metadata describing some containers belonging to kitchen accessories of the Alessi's collection are inserted as instances of the ontology. In particular, a proof of concepts example has been set up considering the objects depicted in Figure 2.

| IceBucket_28 | Jug_26 | WateringCan_1 | Kettles_19 | Jug_24 |
|---|---|---|---|---|
| 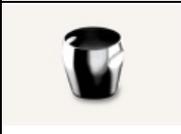 | 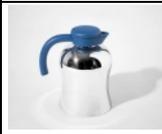 | 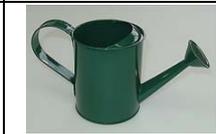 | 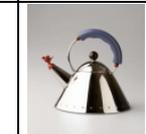 | 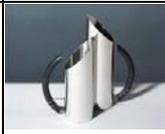 |
|  | MIlkPot_22 | FruitBowl_30 | Kettles_20 | OilCruet_36 |
|  | 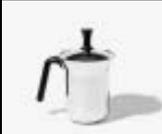 | 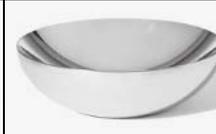 | 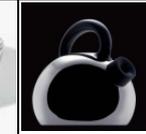 | 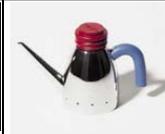 |

**Figure 2: Objects represented as instances in the ontology. All the images except for wateringCan_1 are included thanks to the courtesy of Alessi, http://www.alessi.it.**

### 3.2 Scenario and application context formalization

We consider a scenario where a new product has to be designed for a well defined purpose. The designer needs to browse the company's repository in order to analyze the objects already produced and to get new ideas for its design. In particular, he is aware his company pushes for stylish product having unconventional form, for example, an object whose form does not necessarily recall its real usage.

We assume that the designer could browse the repository according to the following different perspectives:

- First perspective: the user browses the objects analyzing their decomposition in parts to figure out how the objects have been previously realized and to discover commonality among objects of different typologies.
- Second perspective: the user browses the objects analyzing their possible usage to discover commonalities among objects belonging to distinct typologies.

The semantic similarity among instances is worked out to browse the models according to their metadata. Notice that if we considered a context independent semantic similarity the two perspectives would require the definition of two distinct similarity measures. However, applying the context dependent semantic similarity we have proposed in [8][9], it is possible to face with this scenario simply by defining two distinct application contexts.

The definition of the two contexts requires the identification of relevant metadata in the two perspectives and how metadata have to be compared during the similarity assessment. In particular, according to the formalism defined in [8][9] and the ontology specified in Figure 1, the two application context can be defined as follows.

- Application context "Part" for the first perspective. In the first perspective, the designer wants to compare objects in terms of their functional parts. The information that the user needs are spread in the metadata model which is encoded as the ontology in Figure 1. In particular, the entities that the user needs to consider are the categories of the functional parts and the functionalities provided by the parts. Both are compared by an operation of intersection. Such an operation induces that the more two objects have parts with categories and functionalities in common, the more they are similar.

  The application context can be defined by the following function $AC_{Part}$:

  $[Object] \xrightarrow{AC_{Part}} \{\{\},\{(hasPart,Simil)\}\}$

  $[Object.hasPart] \xrightarrow{AC_{Part}} \{\{\},\{(hasPartCategory,Inter),(hasFunctionality,Inter)\}\}$

  The similarity among *Object* instances is recursively assessed in terms of the instances of *FunctionalPart* reachable through the relation *hasPart*. In fact, the operation *Simil* is applied to the relation *hasPart*, moving the focus of similarity assessment from the instances of *Object* to their related instances of *FunctionalPart*. These instances are compared according to the specification defined in the second rows of the function $AC_{Part}$ for the recursion path [Object.hasPart]. Here, it is stated that the similarity assessment among *FunctionalPart* reachable through *hasPart*, is performed by considering the relations *hasPartCategory* and *hasFunctionality*. Both these relations are taken into account through the operation *Inter*, thus the instances similarity will be proportional to the intersection of these relation values.

- Application context "Usage" for the second perspectives. In the second perspective, the designer is interested in comparing objects according to their usage. The entities in the ontology which are relevant for the similarity criteria are: whether or not the object might contain solid, their liquid capacity; the tasks supported by the object. The more two objects share the same tasks and result similar in their liquid capacity and in the ability of containing solid, the greater is their similarity.

The application context can be defined by the following function $AC_{Usage}$:

$$[Object] \xrightarrow{AC_{Usage}} \{\{(mightContainSolid, Simil), (liquidCapacityInLiters, Simil)\}, \{(hasCharacterisingTask, Inter)\}\}$$

According to $AC_{Usage}$ the similarity among *Object* instances is assessed considering the following information: their similarity in ability of containing solid represented through the boolean attribute *mightContainSolid*; the similarity in their liquid capacity encoded as the float attribute *liquidCapacityInLiters;* and the intersection of *Task* instances associated to objects through the relation *hasCharacteringTask*.

### 3.3 Results

The semantic similarity approach is applied to the containers in Figure 2. Their metadata have been inserted in the ontology as instances of the classes *Object* and *FunctionalPart* as illustrated in Table 1 and Table 2.

Similarity is worked out according to the criteria induced by the contexts "Part" and "Usage".

In the following, we first illustrate and analyze the results of the similarity in form of similarity matrix and successively, we discuss the results obtained by adopting the similarity as ranking measure.

Figure 3 illustrates the results: a similarity matrix for each context is obtained. Each row i and each column j of the matrix represents an object in the repository. The grey level of the pixel (i,j) represents the similarity value Sim(i,j) between the two objects located at row i and column j: the darker the colour, the more similar are the two objects. The two matrixes offer an overview of the results useful to draw the following considerations: (i) the two matrixes are rather dissimilar as demonstration that the definition of two contexts induces two different similarity assessments; (ii) both the matrix diagonals are formed by black pixel, as demonstration that each object compared to itself has similarity equal to 1; (iii) both the matrixes are asymmetric then the similarity is asymmetric too.

Further considerations spring to mind considering the assumption pertaining to the concept of containment. For example, let perform a deeper analysis of Figure 3 (a). Taking into account the context "Part" two objects are as similar as they have parts of the same typologies, the "containment" represented as asymmetry induces two objects to have similarity equal to 1 either if they are equal or if the typologies of parts of the first object are contained in the second. However, in the case they are not equal, the similarity between the second and the first is less than 1. For example, the similarity between *FruitBowl_30* and all the other containers is equal to 1 (black pixel) as it is characterized by parts which are in common to all containers. On the contrary, the similarity between the Kettles and the *FruitBoil_30* is very low (clear grey level of the pixel), because kettles are made of many parts not included in *FruitBowl_30.*

Table 3 and Table 4 illustrate the ranking induced by the semantic similarity respectively considering the context "Part" and "Usage". Each row refers to a query object and illustrates the retrieved objects ordered according with the similarity values. The first position shows objects retrieved as the most similar. Each retrieved object is represented by its instance name and the correspondent similarity value.

| InstanceID | MightContain Solid | LiquidCapacity inLiters | Has Characterizing Task | HasPart |
|---|---|---|---|---|
| IceBucket_28 | True | 2.5 | toContain toCool | LiquidProofConcavity_68 SupportingBase_67 |
| Jug_26 | False | 0.8 | toContain toPour | Handle_3 Neck_52 LiquidProofConcavity_50 SupportingBase_51 |
| Jug_24 | False | 0.7 | toContain toPour | CircularNeckToPour_58 LiquidProofConcavity_57 SupportingBase_56 |
| WateringCan_1 | False | 3.0 | toPour | Spout_7 LiquidProofConcavity_5 Handle_8 SupportingBase_34 |
| OilCruet_36 | False | 0.3 | toContain toPour | Handle_23 Cover_24 Spout_22 LiquidProofConcavity_20 SupportingBase_37 |
| Kettles_19 | False | 1.0 | toPour toBoil | Whistle_6 Spout_32 Cover_31 LiquidProofConcavity_29 Handle_30 SupportingBase_39 |
| Kettles_20 | False | 1.0 | toPour toBoil | Whistle_8 Spout_44 Cover_42 LiquidProofConcavity_45 Handle_43 SupportingBase_40 |
| MIlkPot_22 | False | 0.5 | toHeat toBoil toPour | Cover_19 Lip_15 LiquidProofConcavity_17 Handle_16 SupportingBase_36 |
| FruitBowl_30 | True | 3.0 | toContain | LiquidProofConcavity_3 SupportingBase_62 |

**Table 1: Metadata of containers: instances in the class Object.**

| FunctionalPart | hasFunctionality | hasPartCategory |
|---|---|---|
| LiquidProofConcavity_* | ToContain | LiquidProofConcavity |
| SupportingBase_* | ToStabilize | SupportingBase |
| Handle_* | ToLift | Handle |
| Whistle_* | ToWhistle | Whistle |
| Neck_* | ToPour | Neck |
| Spout_* | ToPour | Spout |
| Lip_* | ToPour | Lip |
| CircularNeckToPour_* | ToPour | CircularNeckToPour |
| Cover_* | ToCover | Cover |

**Table 2: Instances of the class FunctionalPart related by hasPart to the objects in Table 1. The columns show respectively the instance name (the number of instance is replaced by the wildcard *), the value for the relation hasFunctionality, and the value for the relation hasPartCategory.**

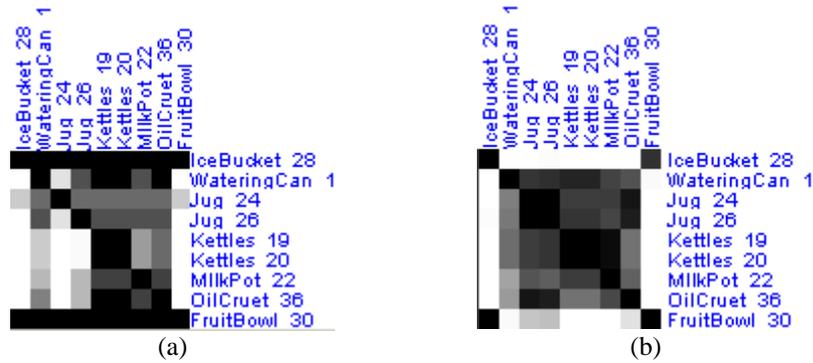

**Figure 3: (a) similarity matrix for context "Part", (b) similarity matrix for context "Usage". The more the instances at row i and column j are similar, the darker is the pixel (i,j).**

Table 3 shows the ranking applying the semantic similarity in the context "Part". Notice that ***IceBucket_28*** and ***FruitBowl_30*** are "contained" in all the other objects, thus all similarities are equal to 1. The containment eases the browsing of objects. For example, it allows to move from ***IceBucket_28*** to the other objects being aware that all the parts characterizing ***IceBucket_28*** still are present in the other objects. Otherwise, if the similarity is less then 1, it means that moving from an object to the other some features are lost. This information could be useful for the designer in order to evaluate the complexity of an object with respect to another. For example, assuming that an object x is contained in an object y, if the designer wants to estimate the increase of complexity moving from x to y, the user can consider the similarity of y with respect to x: the more such similarity is low, the higher is the complexity of the object y with respect to x.

Table 4 shows the object ranking induced by the semantic similarity in the context "Usage". Beside only few instances are adopted in the case study, the results are more than encouraging. It stands up that for each object a reasonable replacement may be selected. ***FruitBowl_30*** is the best replacement for ***IceBucket_28***: although a fruit bowl does not maintain the temperature, ***FruitBowl_30*** is the object which better can act as ice bucket. The kettles are the best replacement for ***WateringCan_1***: beside ***WateringCan_1*** and kettles largely differ in their liquid capacity, the kettles are better to pour than the ***IceBucket_28*** or the ***FruitBowl_30***. Also the jugs have a high ranking to replace ***WateringCan_1***. Looking at the picture in Figure 2, it appears quite reasonable. Of course the ***Jugs_24*** is replaced with ***Jug_26*** and vice versa, but also ***OilCruet_36*** is considered. That is because, according to Table 4, jugs and ***OilCruet_36*** have exactly the same values for the relation ***hasCharacterizingTask***. Kettles are replaced by kettles and ***MilkPot_22***, actually they are the only objects that can boil. Similar considerations can be made for the other results. The similarity induced by the context "Usage" supports in the scenario aforementioned. Let us suppose the designer has to design a new watering can. User browses the existing *WateringCan_1* realizing that the kettles and *WateringCan_1* have similar functionalities. This hint might result inspiring to get an object whose appearance is unconventional, for example it could spring in the designer's mind that a watering-can can look like a big kettle.

| Query Object | | Ranking Position | | | | | | |
|---|---|---|---|---|---|---|---|---|
| | | 1° | 2° | 3° | 4° | 5° | 6° | 7° |
| | IceBucket_28 | ALL | | | | | | |
| | | 1,0000 | | | | | | |
| | WateringCan_1 | Kettles_19 Kettles_20 OilCruet_36 | Jug_26 MIlkPot_22 | Jug_24 | IceBucket_28 FruitBowl_30 | | | |
| | | 1,0000 | 0,8750 | 0,6250 | 0,5000 | | | |
| | Jug_24 | WateringCan_1 Jug_26 Kettles_19 Kettles_20 MIlkPot_22 OilCruet_36 | IceBucket_28 FruitBowl_30 | | | | | |
| | | 0,8333 | 0,6667 | | | | | |
| | Jug_26 | WateringCan_1 Kettles_19 Kettles_20 | MIlkPot_22 OilCruet_36 | Jug_24 | IceBucket_28 FruitBowl_30 | | | |
| | | 0,8750 | 0,8750 | 0,6250 | 0,5000 | | | |
| | Kettles_19 | Kettles_20 | OilCruet_36 | MIlkPot_22 | WateringCan_1 | Jug_26 | Jug_24 | IceBucket_28 FruitBowl_30 |
| | | 1,0000 | 0,8333 | 0,7500 | 0,6667 | 0,5833 | 0,4167 | 0,3333 |
| | Kettles_20 | Kettles_19 | OilCruet_36 | MIlkPot_22 | WateringCan_1 | Jug_26 | Jug_24 | IceBucket_28 FruitBowl_30 |
| | | 1,0000 | 0,8333 | 0,7500 | 0,6667 | 0,5833 | 0,4167 | 0,3333 |
| | MIlkPot_22 | Kettles_19 Kettles_20 OilCruet_36 | WateringCan_1 Jug_26 | Jug_24 | IceBucket_28 FruitBowl_30 | | | |
| | | 0,9000 | 0,7000 | 0,5000 | 0,4000 | | | |
| | OilCruet_36 | Kettles_20 Kettles_19 | MIlkPot_22 | WateringCan_1 | Jug_26 | Jug_24 | IceBucket_28 FruitBowl_30 | |
| | | 1,0000 | 0,9000 | 0,8000 | 0,7000 | 0,5000 | 0,4000 | |
| | FruitBowl_30 | ALL | | | | | | |
| | | 1,0000 | | | | | | |

**Table 3: Object ranking adopting the similarity induced by the context "Part".**

The case study presented in this paper shows that the similarity among objects should consider their semantics as well as its geometry and structure. In fact there are cases where the solely geometry does not characterize completely the object. A trivial example can be found considering the second perspective implemented by the context "Usage". According to such a parameterization of the similarity, *WateringCan_1* results more similar to *MilkPot_22* than to *OilCruet_36*. That is sound because *MilkPot_22* and *WateringCan_1* can be used to boil, whilst the *oilCruet_36* can not. However, if we considered a similarity assessment based on a geometric and structural comparison of the objects as those developed in the field of Shape Modeling (e.g. [21]), **WateringCan_1** would result more similar to *OilCruet_22* than to *MilkPot_22*. That does not mean the geometrical-structural similarity is useless. On the contrary, it is indispensable in many applications. By the way, the need of metadata and ontologies to represent resources is rapidly growing in the community of Computer Graphics as well as Multimedia. Tools to semantically annotate the 3D models are emerging (e.g. Shape annotator [22]). Then, whenever the models are equipped with ontology driven metadata, the use of context dependent semantic similarity is indispensable to support a broader range of applications.

| | | Ranking Position | | | | | | | |
|---|---|---|---|---|---|---|---|---|---|
| | | 1° | 2° | 3° | 4° | 5° | 6° | 7° | 8° |
| Query Object | IceBucket_28 | FruitBowl_30 | Jug_26 | Jug_24 | WateringCan_1 | OilCruet_36 | Kettles_19 Kettles_20 | MilkPot_22 | |
| | | 0,8030 | 0,3283 | 0,3125 | 0,3030 | 0,2381 | 0,1905 | 0,1111 | |
| | WateringCan_1 | Kettles_19 Kettles_20 | Jug_26 | Jug_24 | MilkPot_22 | OilCruet_36 | FruitBowl_30 | IceBucket_28 | |
| | | 0,8333 | 0,8070 | 0,7928 | 0,7619 | 0,7273 | 0,3333 | 0,3030 | |
| | Jug_24 | Jug_26 | OilCruet_36 | MilkPot_22 | Kettles_19 Kettles_20 | WateringCan_1 | IceBucket_28 | FruitBowl_30 | |
| | | 0,9778 | 0,8667 | 0,7778 | 0,7745 | 0,6261 | 0,3125 | 0,2928 | |
| | Jug_26 | Jug_24 | OilCruet_36 | Kettles_19 Kettles_20 | MilkPot_22 | WateringCan_1 | IceBucket_28 | FruitBowl_30 | |
| | | 0,9778 | 0,8485 | 0,7963 | 0,7564 | 0,6404 | 0,3283 | 0,3070 | |
| | Kettles_19 | Kettles_20 | MilkPot_22 | Jug_26 | Jug_24 | WateringCan_1 | OilCruet_36 | IceBucket_28 | FruitBowl_30 |
| | | 1,0000 | 0,8889 | 0,7963 | 0,7745 | 0,6667 | 0,6538 | 0,1905 | 0,1667 |
| | Kettles_20 | Kettles_19 | MilkPot_22 | Jug_26 | Jug_24 | WateringCan_1 | OilCruet_36 | IceBucket_28 | FruitBowl_30 |
| | | 1,0000 | 0,8889 | 0,7963 | 0,7745 | 0,6667 | 0,6538 | 0,1905 | 0,1667 |
| | MilkPot_22 | Kettles_19 Kettles_20 | Jug_24 | Jug_26 | OilCruet_36 | WateringCan_1 | IceBucket_28 | FruitBowl_30 | |
| | | 0,7778 | 0,7222 | 0,7009 | 0,6944 | 0,5397 | 0,1111 | 0,0952 | |
| | OilCruet_36 | Jug_24 | Jug_26 | MilkPot_22 | Kettles_19 Kettles_20 | WateringCan_1 | IceBucket_28 | FruitBowl_30 | |
| | | 0,8667 | 0,8485 | 0,7500 | 0,6538 | 0,5606 | 0,2381 | 0,2273 | |
| | FruitBowl_30 | IceBucket_28 | Jug_26 | Jug_24 | OilCruet_36 | WateringCan_1 | Kettles_19 Kettles_20 | MilkPot_22 | |
| | | 0,9697 | 0,4737 | 0,4595 | 0,3939 | 0,3333 | 0,1667 | 0,0952 | |

**Table 4: Object ranking adopting the similarity induced by the context "Usage".**

## 4 Conclusion and future work

The paper illustrates the importance of adopting a context dependent semantic similarity approach to browse multimedia contents. It is exemplified with a case study for industrial design to support a designer-oriented browsing of a repository of 3D objects. Our semantic similarity takes into account the influence of the application context and provides an asymmetric measure to point out the concept of "containment". Even if the ontology considered is quite simple, it is sufficient to demonstrate the usefulness of our approach.

The similarity dependence from the context is mandatory in order to support different designer perspectives, whereas the "containment" in terms of similarity asymmetry results to be a sophisticated tool for browsing. Moreover, the case study shows examples where the semantic-based approach is able to recognize similarities that can not be identified by geometric/structural based approaches.

Future works concern research activities at diverse levels. Advance for 3D search engine will be investigated integrating our approach with the existing geometric and structural similarity measures. In particular, limits and benefits of both approaches should be pointed out in order to fruitfully combine them in the next wave of 3D search engines. We also foresee specific advances in the domain of industrial design by improving the ontology and considering larger repository. Finally, other case studies on other kinds of multimedia contents can be investigated.

# 5 Acknowledgement

This work is partially supported by the EU Network of Excellence AIM@SHAPE.